\documentclass[12pt,prd,amsmath,amssymb,nofootinbib,floatfix,superscriptaddress,tightenlines,preprint]{revtex4}

\usepackage{epsfig,latexsym,cancel,amssymb,amsmath,verbatim,mathrsfs}
\usepackage{color}
\usepackage{graphicx}
\usepackage{hyperref,dcolumn,bm,subfigure}
\usepackage{collref}

\newcommand{\be}{\begin{equation}}
\newcommand{\ee}{\end{equation}}
\newcommand{\bea}{\begin{eqnarray}}
\newcommand{\eea}{\end{eqnarray}}
\newcommand{\ba}{\begin{array}}
\newcommand{\ea}{\end{array}}

\long\def\symbolfootnote[#1]#2{\begingroup%
\def\thefootnote{\fnsymbol{footnote}}\footnote[#1]{#2}\endgroup}

\newcommand{\beq}{\begin{equation}}
\newcommand{\eeq}{\end{equation}}

\newcommand{\tev}{\, {\rm TeV}}
\newcommand{\gev}{\, {\rm GeV}}

\newcommand{\cms}{{\rm cm}^3/{\rm s}}
\newcommand{\AFB}{A_{FB}}
\newcommand{\AC}{A_{C}}
\newcommand{\Mtt}{M_{t\bar{t}}}
\newcommand{\fbinv}{\rm fb^{-1}}

\begin{document}

\title{Top-flavored dark matter and the forward-backward asymmetry} %

\author{Abhishek Kumar} 
\email{abhishek@triumf.ca}
\affiliation{Theory Group, TRIUMF, 4004 Wesbrook Mall, Vancouver, BC, 
V6T 2A3, Canada}
\author{Sean Tulin} 
\email{tulin@umich.edu}
\affiliation{Department of Physics, University of Michigan, Ann Arbor, MI 48109}

\date{\today}

\preprint{MCTP-13-04}

\begin{abstract}

We propose a simple model where dark matter (DM) carries top flavor and couples to the Standard Model through the top quark within a framework of minimal flavor violation (MFV).  Top-flavored DM can explain the anomalous top forward-backward asymmetry observed at the Tevatron, while remaining consistent with other top observables at colliders.  By virtue of its large coupling to top, DM acquires a sizable loop coupling to the $Z$ boson, and the relic density is set by annihilation through the $Z$.  We also discuss contraints from current direct detection searches, emphasizing the role of spin-dependent searches to probe this scenario.

\end{abstract}

\maketitle

\section{Introduction}\label{intro}

The nature of dark matter (DM) remains mysterious.  Aside from its gravitational influence, the particle physics properties of DM are largely unknown, and therefore it is worthwhile to explore different avenues for how DM may couple to the Standard Model (SM).  One interesting possibility is that the dominant couplings of DM to the SM arise through the top quark.  As the heaviest known elementary particle, the top quark plays a fundamental role in many extensions of the SM, and experimental studies of the top quark
are vital to validate our understanding of the weak scale.  
Anomalies in the top sector could be harbingers of physics beyond the SM, and this new physics may be connected to DM.

At the Tevatron, both CDF and D0 collaborations have measured an anomalously large top forward-backward asymmetry ($\AFB$).  The measured parton-level inclusive $\AFB$ in the $t\bar t$ rest frame is (after unfolding and background subtraction)~\cite{Aaltonen:2012it,Abazov:2011rq}
\be \label{AFBexpt}
\AFB = \left\{ \begin{array}{ll} 0.164 \pm 0.045 & \;\;{\rm CDF \; (9.4 \; \fbinv)}
\\ 0.196 \pm 0.065 & \;\;{\rm D0 \; (5.4 \; \fbinv)} \end{array} \right. 
\ee
for semileptonic $t\bar t$ events, in excess over the SM value $A_{FB}^{\rm SM} = 0.088 \pm 0.006$~\cite{Bernreuther:2012sx} by $\sim 2 \sigma$.
Notably, the discrepancy is larger for higher $t\bar t$ invariant mass ($\Mtt$).  For the high invariant mass bin ($\Mtt > 450$ GeV), CDF has found $\AFB^{\rm high} = 0.295 \pm 0.066$~\cite{Aaltonen:2012it}, a $\sim 2.5 \sigma$ deviation from the SM value $A_{FB}^{\rm high, SM} = 0.129^{+ 0.008}_{-0.006}$.  CDF has also observed a sizable $\AFB$ in dileptonic $t\bar t$ events, albeit with larger errors~\cite{CDF-Note-10436}.  However, other top observables at both the Tevatron and LHC appear consistent with the SM thus far.

In this work, we propose a simple model where DM, denoted $\chi_t$, carries top flavor and couples directly to the top quark.  We take $\chi_t$ to be the lightest component of a multiplet $\chi$ carrying SM quark flavor, and we couple $\chi$ to quarks according to the principle of minimal flavor violation (MFV)~\cite{Chivukula:1987py,Hall:1990ac,Buras:2000dm,D'Ambrosio:2002ex} through a scalar mediator $\phi$ slightly heavier than $t$.  DM stability is ensured automatically by MFV with no additional assumptions~\cite{Batell:2011tc}.  Although our model is phenomenological, we note that one realization of top-flavored DM is provided by warped grand unified models, where DM is a sterile neutrino partner of the top~\cite{Agashe:2004ci,Agashe:2004bm,Belanger:2007dx}.  (We note other related works exploring top-DM interactions~\cite{Jackson:2009kg,Cheung:2010zf,Zhang:2012da,Haisch:2012kf} and flavored DM~\cite{MarchRussell:2009aq,Kile:2011mn,Agrawal:2011ze}.)

Top-flavored DM (TFDM) provides a number of interesting implications for collider phenomenology.   Our model can explain $\AFB$, while being consistent with other top observables.  To generate $A_{FB}$, an $\mathcal{O}(1)$ forward-backward asymmetry is generated in $u \bar u \to \phi \phi^*$, which is converted into an asymmetry in $t \bar t$ through decays $\phi \to t \bar\chi_t$.  This mechanism was proposed for a supersymmetry-inspired model in ref.~\cite{Isidori:2011dp}, but was later shown not to provide a viable DM candidate~\cite{Hektor:2011ms}.  However, TFDM is a viable candidate for DM, evading many constraints associated with the model in ref.~\cite{Isidori:2011dp} as a natural consequence of MFV. Additionally, we study the collider implications of TFDM for the LHC, in particular the top charge asymmetry ($A_C$) and  top-jet resonance searches~\cite{Gresham:2011dg}. 

On the cosmological side, TFDM can explain the observed DM relic abundance.  As we show, DM acquires a one-loop coupling to the $Z$ boson that is enhanced by the top mass.  DM freeze out in the early Universe is governed by annihilation through the $Z$ to SM fermions, providing the correct relic density for the same parameter region that can account for top $A_{FB}$.  TFDM in the local halo can give observable signals in direct detection experiments.  In principle, the spin-independent (SI) DM-nucleon coupling can be large enough to explain the positive signals by CoGeNT~\cite{Aalseth:2010vx} and CRESST-II~\cite{Angloher:2011uu}, however there is strong tension with constraints from XENON~\cite{Aprile:2012nq,Angle:2011th} and CDMS~\cite{Ahmed:2009zw,Ahmed:2010wy} experiments.  While SI signals may be highly suppressed (through a Majorana splitting), TFDM predicts an irreducible limit on the spin-dependent (SD) cross section that cannot be evaded without additional channels to set the relic density. 

TFDM provides a minimal model to confront possible new physics in the top sector and DM simultaneously.  Moreover, these different considerations {\it independently} point toward the same mass range for DM: $m_{\chi_t} \sim 20- 90$ GeV.
In the remainder of this paper, we discuss each of these topics in detail.  In Sec.~\ref{sec:model}, we present the TFDM model.  We study top-related collider signatures at the Tevatron and LHC (in particular $\AFB$) in Sec.~\ref{sec:collider}.
Relic density, direct detection, and other constraints are discussed in Sec.~\ref{sec:det}.  Finally, we conclude in Sec.~\ref{sec:conclude}.

\section{Model}
\label{sec:model}

The principle of MFV ensures that new physics entering at the weak scale preserves the successful SM predictions of precision flavor measurements, which are sensitive to much higher mass scales~\cite{Isidori:2010kg}.  The quark sector of the SM possesses a global flavor symmetry under ${G}_{q} = SU(3)_{Q_L} \times SU(3)_{u_R} \times SU(3)_{d_R}$ that is broken only by Yukawa interactions.\footnote{We omit the lepton flavor symmetry $SU(3)_{\ell_L} \times SU(3)_{e_R}$, which, though unimportant in our model, is relevant for lepton-flavored DM~\cite{MarchRussell:2009aq,Agrawal:2011ze}.}  Under MFV, it is assumed that new physics also preserves the symmetry $G_q$, up to insertions of the Yukawa matrices $Y_{u,d}$ acting as spurion fields.  Although in most DM models, the DM particle is a singlet under $G_q$, it is an interesting and less-explored possibility that DM transforms nontrivially under $G_q$.  One appealing feature of flavored DM is that DM stability is imposed automatically by the flavor symmetry~\cite{Batell:2011tc} (for most but not all representations of $G_q$), whereas other DM models typically require additional symmetry assumptions for stability.

In our model, we introduce two new fields, a Dirac fermion $\chi$ and a complex scalar $\phi$:
\be
\chi \sim  (\mathbf{1},\mathbf{1},0)_{\rm SM} \times  (\mathbf{1}, \mathbf 3, \mathbf{1})_{G_q} \, , \quad \phi \sim  (\mathbf{3},\mathbf{1},2/3)_{\rm SM} \times  (\mathbf{1}, \mathbf 1, \mathbf{1})_{G_q} \, ,
\ee
where the two sets of numbers indicate quantum numbers under the SM gauge group $SU(3)_C \times SU(2)_L \times U(1)_Y$ and $G_q$, respectively.
That is, $\chi$ is a gauge-singlet, flavor multiplet comprised of three states $\chi = (\chi_u, \chi_c, \chi_t)$, while $\phi$ is a flavor-singlet, color-triplet scalar with electric charge $+2/3$ which mediates interactions between $\chi$ and quarks.  The Yukawa spurions transform as $Y_u \sim (\mathbf 3,\bar{\mathbf{3}},\mathbf 1)$ and $Y_d \sim (\mathbf 3,\mathbf 1,\bar{\mathbf{3}})$ under $G_{q}$.  

The spectrum for the $\chi$ states is constrained by MFV.  Keeping only the top Yukawa $y_t$, the mass term can be written as an expansion in powers $(Y_u^\dagger Y_u)^n$ with coefficients $m_n$:
\be
- \mathscr{L}_{\rm mass} =  \bar \chi (m_0 + m_1Y_u^\dagger Y_u  + ... \, ) \chi = m_{\chi_u} \bar{\chi}_u \chi_u + m_{\chi_c} \bar{\chi}_c \chi_c  + m_{\chi_t} \bar{\chi}_t \chi_t \; ,
\ee
where the ellipsis denotes higher-order powers.  Thus, $\chi_u$ and $\chi_c$ are (approximately) degenerate, with mass determined by the zeroth-order term ($m_{\chi_u} \approx m_{\chi_c} \approx m_0$), while the top-flavored state $\chi_t$ has a different mass $m_{\chi_t} \!=\! (m_0 + m_1 y_t^2 + ...) $ from Yukawa insertions.  Assuming these corrections are large and negative, we take $\chi_t$ to be the lightest state, with mass $m_{\chi_t} \ll m_{\chi_{u}}$.

The scalar mediator $\phi$ couples $\chi$ to up-type quarks $q_R = (u,c,t)_R$.  As above, these couplings can be expressed in powers $(Y_u^\dagger Y_u)^n$ with coefficients $g_n$:
\begin{align} \label{int}
\mathscr{L}_{\textrm{int}} &= \bar{q}_{R} (g_0 + g_1 Y_u^\dagger Y_u + ... \, )  \chi  \phi + \textrm{h.c.} 
=  g_u  \bar{u}_{R} \chi_{u} \phi +g_c  \bar{c}_R \chi_c  \phi + g_t  \bar{t}_{R} \chi_{t}  \phi + \textrm{h.c.}
\end{align}
Thus, according to MFV, $g_u \approx g_c \approx g_0$ is given by the leading term, while $g_t = (g_0 + g_1 y_t^2 + ... )$ can be different due to higher-order insertions.  We take these couplings to be real.

The states $\chi_u$ and $\phi$, as well as the DM state $\chi_t$, play an important role in the phenomenology of our model.  To summarize, the main parameters are the $\chi$ masses ($m_{\chi_u}, m_{\chi_t}$) and couplings $(g_u,g_t)$ to quarks, as well as the scalar mediator mass $m_\phi$.  

Next, we consider a possible Majorana mass term for $\chi_t$:
\be \label{majoranamass}
\mathscr{L}_{\rm Majorana} = \frac{\Delta m}{2} \bar{\chi}_t^C \chi_t + {\rm h.c.} 
\ee
This term violates $G_q$ and cannot be accommodated within MFV, and therefore $\chi_t$ must be Dirac.  Effectively, if we regard the Majorana mass $\Delta m$ as a spurion, its representation is $\Delta m \sim (\mathbf{1},\bar{\mathbf{6}},\mathbf{1})$, which cannot be composed of $Y_{u,d}$.  Nevertheless, it is useful to consider that small deviations from MFV may arise, and with $\Delta m \ne 0$, the single Dirac state $\chi_t$ is split into two Majorana components $\chi_{1,2}$, with mass $m_{1,2} = m_{\chi_t} \pm \Delta m$.  In this case, even a small Majorana term $\Delta m \ll m_{\chi_t}$ becomes important for direct detection phenomenology, discussed in Sec.~\ref{sec:det}.  Clearly, we assume that deviations from MFV do not destabilize DM; this depends on underlying model-building assumptions that are beyond the scope of our phenomenologically-motivated study.

Lastly, we discuss how the model of ref.~\cite{Isidori:2011dp} differs from TFDM.  To explain $\AFB$, ref.~\cite{Isidori:2011dp} introduces a scalar top (``stop'') partner $\tilde t$ (playing the role of $\phi$) and a single light neutralino $\chi^0$ (playing the role of $\chi_u$ and $\chi_t$).  $\AFB$ arises by generating a forward-backward asymmetry in $\tilde t \tilde t^*$, which is converted to $t \bar t$ through decays.  However, having a single light state $\chi^0$ coupled to both $\bar u_R \tilde t$ and $\bar t_R \tilde t$ allows for two decay channels $\tilde t \to t \chi^0$ and $\tilde t \to u \chi^0$; the latter channel is phase space enhanced and dilutes $\AFB$ unless $g_t \gg g_u \sim 1$.  (Also, the charm interaction must be very suppressed due to $D^0$-$\bar D^0$ mixing bounds, $g_c/g_u < 0.06$~\cite{Isidori:2011dp}).  This model is strongly constrained by LHC searches for jets plus missing energy ($E_T^{\rm miss}$)~\cite{:2012mfa,:2012rz} via pair production $\tilde t \tilde t^* \to u \bar u \chi^0 \chi^0$, and monojets plus $E_T^{\rm miss}$ \cite{Chatrchyan:2012me,ATLAS:2012ky} via $ug \to \tilde{t} \chi^0 \to u \chi^0 \chi^0$.  Also, identifying $\chi^0$ as DM is excluded by direct detection bounds, due to the large tree-level $\chi^0$-$u$ coupling~\cite{Hektor:2011ms}.  For TFDM, all these issues are easily avoided by having $\chi$ as a flavor multiplet.  For $m_{\chi_u} > m_\phi$, decays $\phi \to u \bar \chi_u$ and $\phi \to c \bar{\chi}_c$ are kinematically blocked and do not dilute $\AFB$ since $BR(\phi \to t \bar\chi_t) = 1$ without nonperturbatively large coupling $g_t$.  Signals from (mono)jets plus $E_T^{\rm miss}$ are eliminated at tree-level since $\phi \to u \bar\chi_u$ is forbidden (although loop-induced monojet signals can arise~\cite{Haisch:2012kf}). Same-sign top production is not allowed by the Dirac nature of $\chi_u$, and no fine-tuning in the charm sector is required.  Direct detection constraints are weakened since the DM $\chi_t$ couples to $u$ at one-loop (see Sec.~\ref{sec:det}).

\section{Top phenomenology at colliders}
\label{sec:collider}

\subsection{Top forward-backward asymmetry}

In the SM, the $t \bar t$ forward-backward asymmetry arises at order $\alpha_{s}^{3}$ in the cross section, and therefore new physics in the top sector can provide a significant enhancement to $\AFB$.  Several models have been proposed to explain the top asymmetry ({e.g.} axigluons~\cite{Frampton:2009rk, Cao:2010zb, Alvarez:2011hi, Hewett:2011wz, Bai:2011ed, Tavares:2011zg, Gross:2012bz, Cvetic:2012kv} and flavor-changing scalar or vector bosons~\cite{Jung:2009jz, Shu:2009xf, Arhrib:2009hu, Dorsner:2009mq, Djouadi:2009nb, Ko:2011di, Jung:2011zv, Blum:2011fa, Jung:2011id, Nelson:2011us, Shelton:2011hq, AguilarSaavedra:2011ug, Grinstein:2011yv, AguilarSaavedra:2011ci, Cao:2011hr, Berger:2011ua, Bhattacherjee:2011nr, AguilarSaavedra:2011zy}), mostly relying on interference between new physics (NP) and gluon amplitudes in $t\bar t$ production.  However, the simplest models are strongly constrained by top observables at the Tevatron and LHC~\cite{Gresham:2011pa,Gresham:2011fx, Atwood:2013xg, Biswal:2012mr, Gross:2012bz, Gresham:2012kv}, as well as by atomic parity violation constraints~\cite{Gresham:2012wc}, requiring additional model-building ingredients to be viable.  

\begin{figure}[t!]
\centering
\subfigure[]{
\includegraphics[width=0.45\textwidth,viewport={46 521 515 777}]{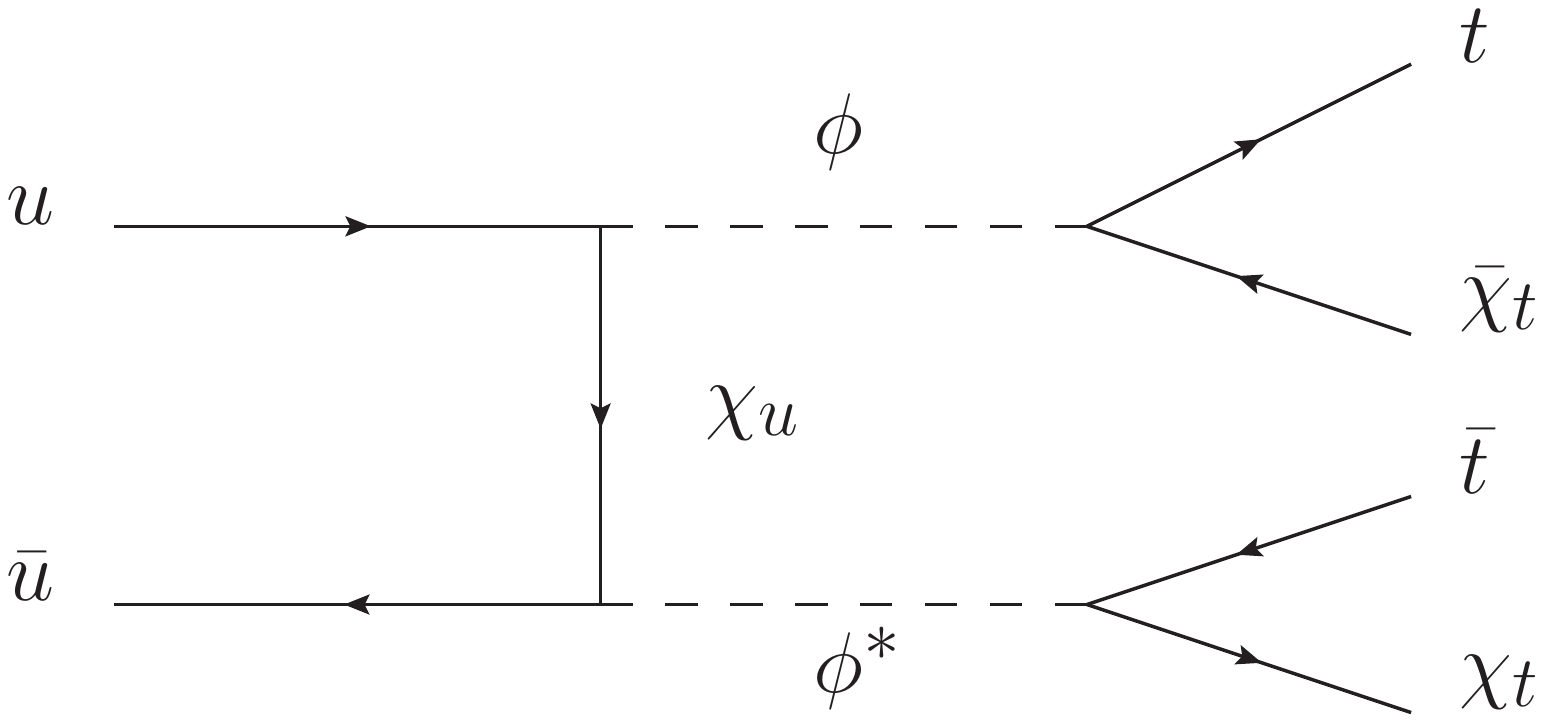}
\label{fig:AFBfig}   
}
\qquad 
\subfigure[]{
\includegraphics[width=0.35\textwidth,viewport={100 380 470 780}]{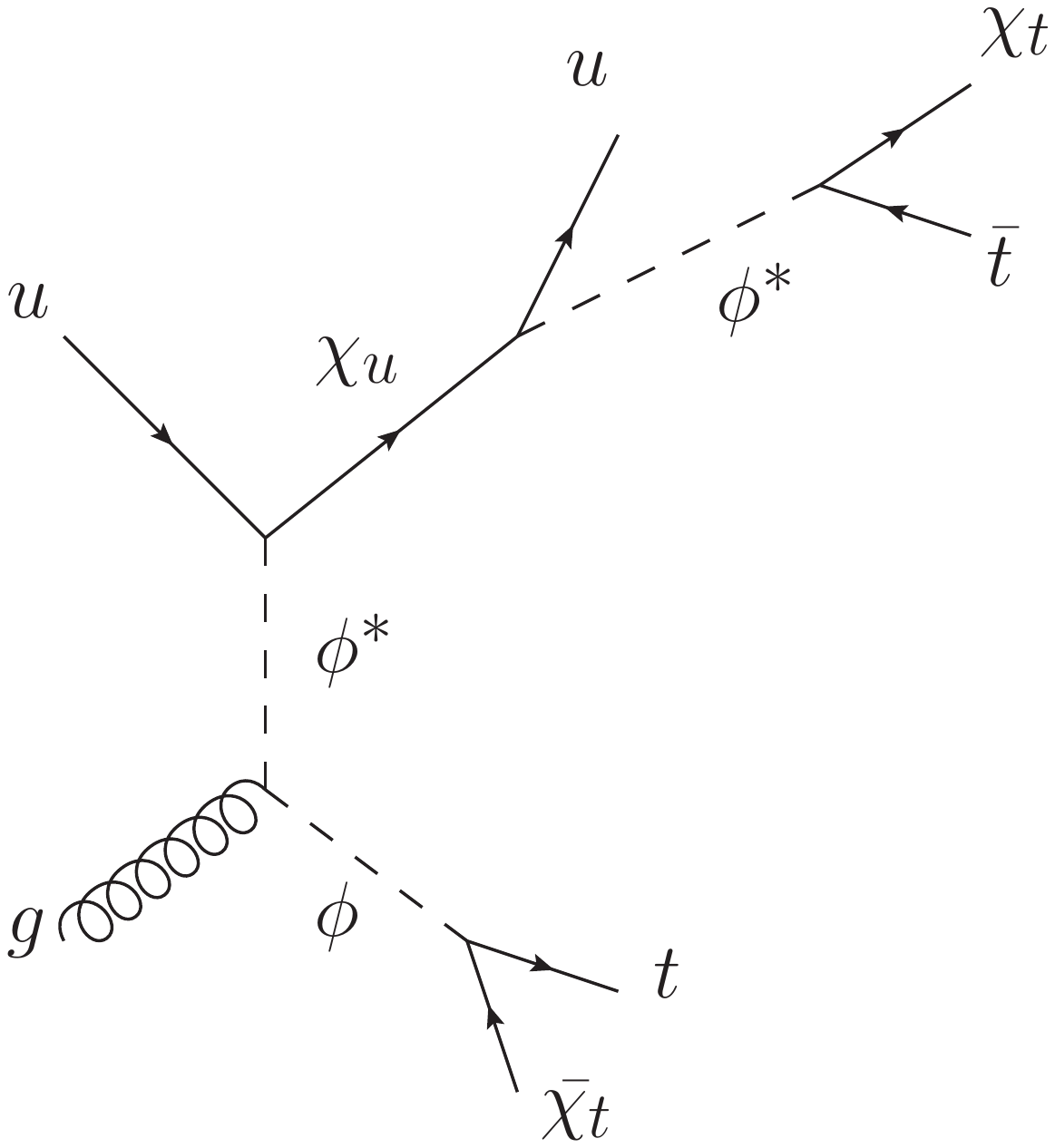}
\label{fig:ACneg}   
}
\caption{(a): $\AFB$ is generated via $\phi$ production through a $t$-channel mediator (b): The dominant contribution to a negative $A_{C}$}  \label{feyn}
\end{figure}

For TFDM, $\AFB$ is generated entirely through NP without interference with the QCD amplitude, shown in Fig.~\ref{fig:AFBfig}, through the mechanism of ref.~\cite{Isidori:2011dp}.  First, an $\mathcal{O}(1)$ forward-backward asymmetry is generated in $\phi \phi^*$ through $t$-channel $\chi_u$ exchange, and then this asymmetry is transmitted to $t\bar t$ through decays $\phi \to t \bar\chi_t$ and $\phi^* \to \bar t \chi_t$.  Fig.~\ref{fig:AFBfig} gives a parton-level differential cross section
\be
\frac{d \hat \sigma(u \bar u \to \phi \phi^*)}{d \cos\theta} =
 \frac{\beta_\phi}{32 \pi \hat s}  \, \frac{ g_u^4\beta_\phi^2 (1-\cos^2\theta)}{(1 + \beta^2_\phi- 2 \beta_\phi \cos\theta  + 4 m_{\chi_u}^2/\hat s)^2} \, , \label{dsigma1}
\ee
where $\beta_\phi = \sqrt{ 1 - 4 m_\phi^2/\hat s}$, which is strongly peaked in the forward direction ($\cos\theta >0$) as long as $4 m_{\chi_u}^2/\hat s \lesssim 1$.  Also, since QCD production of scalars $\phi \phi^*$ is $p$-wave suppressed, this channel does not give a large contribution to inclusive $t\bar{t}$ production.  Searches for $t \bar t$ plus $E_T^{\rm miss}$ (see e.g.~\cite{Aaltonen:2011rr,Aaltonen:2011na,Aad:2012uu}) are evaded by assuming $\phi$ to be nearly degenerate with $t \bar \chi_t$ ({i.e.}, $m_\phi -m_t - m_{\chi_t} \approx 0$).  We also take $m_{\chi_u} > m_\phi$ to maximize $\AFB$ by having ${\rm BR}(\phi \to t \bar\chi_t) = 1$ (since $\phi \to u \bar\chi_u,  c \bar\chi_c$ are forbidden).

To explain the $\AFB$ data, the TFDM model must generate a large inclusive asymmetry, as well as an asymmetry that rises with $\Mtt$, without introducing large corrections to the total inclusive $t\bar{t}$ cross section ($\sigma_{t\bar{t}}$) or the differential cross section ($d\sigma/d\Mtt$).  The inclusive $t\bar{t}$ production cross section is $\sigma_{t\bar{t}} = 7.65 \pm 0.42$ pb \cite{CDF-Note-10926,D0-Note-6363,Schilling:2013nca}, consistent with the theoretical SM predictions of $\sigma_{t\bar t}^{\rm SM} = 7.067 \pm 0.26$ pb \cite{Baernreuther:2012ws}, allowing for $\sigma_{t\bar t}^{\rm NP} = (0.58 \pm 0.49)$ pb at $1\sigma$.

For $d \sigma/d \Mtt$, we quantify this constraint in terms of the $t\bar t$ cross section for $\Mtt > 450$ GeV, measured by CDF to be $\sigma_{t \bar t}^{\rm high} = 1.92 \pm 0.48$ pb~\cite{Aaltonen:2009iz}, while the SM prediction is $\sigma_{t \bar t}^{\rm high, SM} = 2.17 \pm 0.10$ pb~\cite{Ahrens:2010zv}.  This observable, as we show, provides the strongest constaint on our model.  We have also investigated the high $\Mtt$ tail, quantified in terms of the second-highest invariant mass bin ($\Mtt = 700 - 800\,\gev$) following refs.~\cite{Blum:2011up,Blum:2011fa}.  However, within our model, this observable is not as constraining as $\sigma_{t \bar t}^{\rm high}$.

To simulate this signal, we generate NP events ($p\bar{p} \rightarrow t\bar{t}\chi_{t}\bar{\chi}_{t}$) at the partonic level in MadGraph5 v1.4.3 \cite{Alwall:2011uj}, with a $p\bar{p}$ initial state at $\sqrt{s}=1.96\,\tev$ and $m_t = 173.3$ GeV~\cite{Deliot:2013hc}.  We compute the total (SM + NP) $\AFB$ as
\be
\AFB = \AFB^{\rm SM} \, \frac{\sigma_{t\bar t}^{\rm SM}}{\sigma_{t\bar t}^{\rm SM+NP}} + \AFB^{\rm NP} \,\frac{\sigma_{t\bar t}^{\rm NP}}{\sigma_{t\bar t}^{\rm SM+NP}}  \, ,
\ee
where $\AFB^{\rm NP}$ is the asymmetry generated in $t \bar t$ produced from {\it NP-only}.  We work to leading-order (LO) in computing NP cross sections; we take SM quantities given at their highest-available order.  We take $\AFB^{\rm SM}$ to be the most recent next-to-LO (NLO) QCD prediction including LO electroweak contributions~\cite{Hollik:2011ps,Kuhn:2011ri,Manohar:2012rs,Bernreuther:2012sx}.  We expect that the inclusion of (unknown) NLO corrections to NP quantities would likely reduce the allowed parameter space, and therefore our treatment is conservative in the sense of allowing for the largest possible parametric region.

\subsection{Top charge asymmetry}

At the LHC, a related observable is the top charge asymmetry ($A_C$) with respect to the boost direction of $t\bar t$ pair.  The ATLAS and CMS collaborations have measured inclusive asymmetries (at 7 TeV)~\cite{ATLAS:2012an,Chatrchyan:2012cxa}
\be \label{ACexpt}
\AC = \left\{ \begin{array}{ll}- 0.018 \pm 0.036 & \;\;{\rm ATLAS \; (1.04 \; \fbinv)}
\\  0.004 \pm 0.015 & \;\;{\rm CMS \; (5.0 \; \fbinv)} \end{array} \right. \, ,
\ee 
consistent with the SM prediction, $\AC^{\rm SM}=0.0123 \pm 0.0005$ \cite{Kuhn:2011ri,Bernreuther:2012sx}. Additionally, the combined inclusive $t\bar t$ cross section at the LHC is $\sigma_{t\bar{t}} = 173.3 \pm 10.1$ pb \cite{ATLAS:2012dpa}, compared to the NNLL SM calculation, $\sigma_{t\bar{t}}^{\rm SM} = 162.6 \pm 17.1$ pb \cite{Beneke:2011mq} for $m_{t}=173.3$ GeV.

There are two contributions to the charge asymmetry in our model, shown in Fig.~\ref{feyn}, and they have opposite sign. The positive contribution comes from the process $u \bar u \rightarrow \phi \phi^* \to t\bar{t}\chi_{t}\bar{\chi_{t}}$. The negative contribution arises from $u g \rightarrow \chi_{u} \phi$, which in turn decay to $\chi_{u} \rightarrow u \bar{t} \chi_{t}$ and $\phi \rightarrow t \bar{\chi_{t}}$. The incoming $u$-quark is harder than the incoming $g$ at the LHC. As a result, the outgoing $\bar{t}$ is harder than the outgoing $t$, resulting in a negative $A_{C}$. The cross section for the CP conjugate process is an order of magnitude smaller, due to the incoming $\bar{u}$ parton distribution function. As a result, our model can generate a positive $\AFB$ at the Tevatron while also generating $A_{C}$ of either sign at the LHC.  In particular, the process mediated by $\chi_{u} \phi$ generates a sizable $A_{C} \sim -20\%$. This effect, first discussed in \cite{Alvarez:2012ca,Drobnak:2012rb}, can be understood from the parton-level cross section computed from Fig.~\ref{fig:ACneg}
\be
\frac{ d \hat{\sigma}(u g \to \phi \chi_u)}{d \cos\theta} \approx \frac{\beta_\phi}{32 \pi \hat s} \, \frac{g_u^2 g_s^2}{24} \big(  1 - \beta_\phi \cos\theta \big) \; ,
\ee
neglecting terms of order $(m_{\chi_u}^2 - m_\phi^2)/\hat s$.  This process tends to produce $\phi$ anti-aligned with the initial $u$, giving a negative contribution to $A_C$.

To compute the charge asymmetry in our model, we generate the above two NP processes at the partonic level in MadGraph5 v1.4.3, with a $pp$ initial state at $\sqrt{s}=7 \,\tev$. 
Similar to $\AFB$, we compute
\be
\AC = \AC^{\rm SM} \, \frac{\sigma_{t\bar t}^{\rm SM}}{\sigma_{t\bar t}^{\rm SM+NP}} + \AC^{\rm NP} \,\frac{\sigma_{t\bar t}^{\rm NP}}{\sigma_{t\bar t}^{\rm SM+NP}},
\ee
where $\AC^{\rm NP}$ is the asymmetry generated in $t \bar t$ produced from {\it NP-only}.

\subsection{Top+jet resonance}

The heavy $\chi_{u}$ can be searched for in top+jet resonance searches at colliders~\cite{Gresham:2011dg,Aaltonen:2012qn,Aad:2012em}.    This signal arises in $t\bar t j$ events in $t$-channel mediator models; the mediator is produced on-shell, in association with a $t$, and has a two-body decay to $\bar t j$.  Within our model, a top+jet resonance arises from sequential 2-body decays $\chi_u \to u \phi \to u \bar t \chi_t$, where $\chi_t$ is undetected. The top+jet invariant mass is given by
\be
m_{tj}^2 = m_t^2 + \frac{m_{\chi_u}^2 - m_\phi^2}{2 m_\phi^2} \left( m_\phi^2 + m_t^2 - m_{\chi_t}^2 
- \cos\theta \sqrt{ (m_\phi^2 - m_t^2 - m_{\chi_t}^2 )^2 - 4 m_t^2 m_{\chi_t}^2 } \,\right) \; ,
\ee
where $\theta$ is the angle between the $\bar t$ and $u$ in the $\phi$ rest frame.  Since $\phi$ decays isotropically, the top+jet resonance has a ``box'' feature.  However, since collider constraints prefer a squeezed spectrum (such that $m_{\chi_u} - m_{\phi} \ll m_{\chi_u}$ and $m_\phi - m_t - m_{\chi_t} \approx 0$), the width of the box is typically smaller than the jet energy resolution~\cite{ATLAS:1999uwa}.  In this case, we can approximate $m_{t j} \approx m_{\chi_u} - m_{\chi_t}$, which is valid within 5\% over the parameter space of interest.

The ATLAS search \cite{Aad:2012em} places the strongest limits on the production of top+jet resonances within the context of specific models.  For the color-singlet vector model ($W^\prime$), the parton-level cross section is constrained to be less than $(23,14,7)$ pb for $m_{W^\prime} = (200,300,400) \, \gev$, respectively.  Assuming the acceptances and efficiencies are comparable, we apply these limits to TFDM, with the replacement $m_{W^\prime} \to m_{\chi_u} - m_{\chi_t}$, and we extrapolate between these different mass points following \cite{Aad:2012em}.  We do not extrapolate below 200 GeV, since searches~\cite{Aaltonen:2012qn,Aad:2012em} have not been performed in this range.

\subsection{Analysis}

In this section, we identify what is the region of parameter space where TFDM can account for an anomalous $A_{FB}$, while remaining consistent with other observables.  We simulate the NP contribution to top observables in MadGraph.  We emphasize that the top asymmetry generated in TFDM does not necessitate a light $\chi_{t}$, unlike the case in \cite{Isidori:2011dp}. By restricting to a degenerate spectrum, $m_\phi - m_t - m_{\chi_t} \approx 0$, we can increase the mass of both $\chi_{t}$ and $\phi$ while minimizing $E_T^{\rm miss}$ in $t\bar{t}$ production. 

\begin{figure}[h!]
\centering
\subfigure[]{ 
\includegraphics[width=0.47\textwidth]{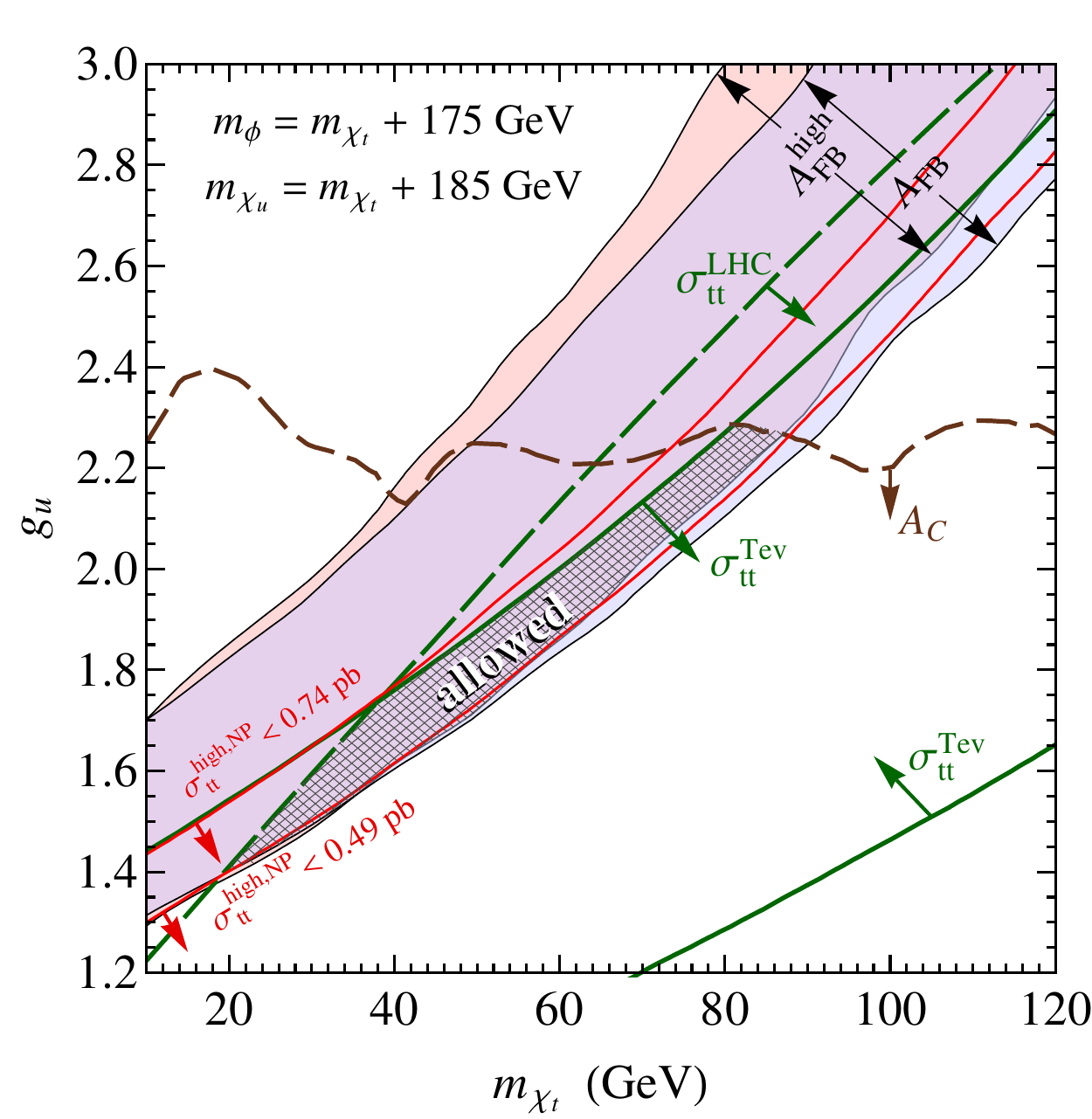}
\label{fig:chitrange}
}
\centering
\subfigure[]{ 
\includegraphics[width=0.47\textwidth]{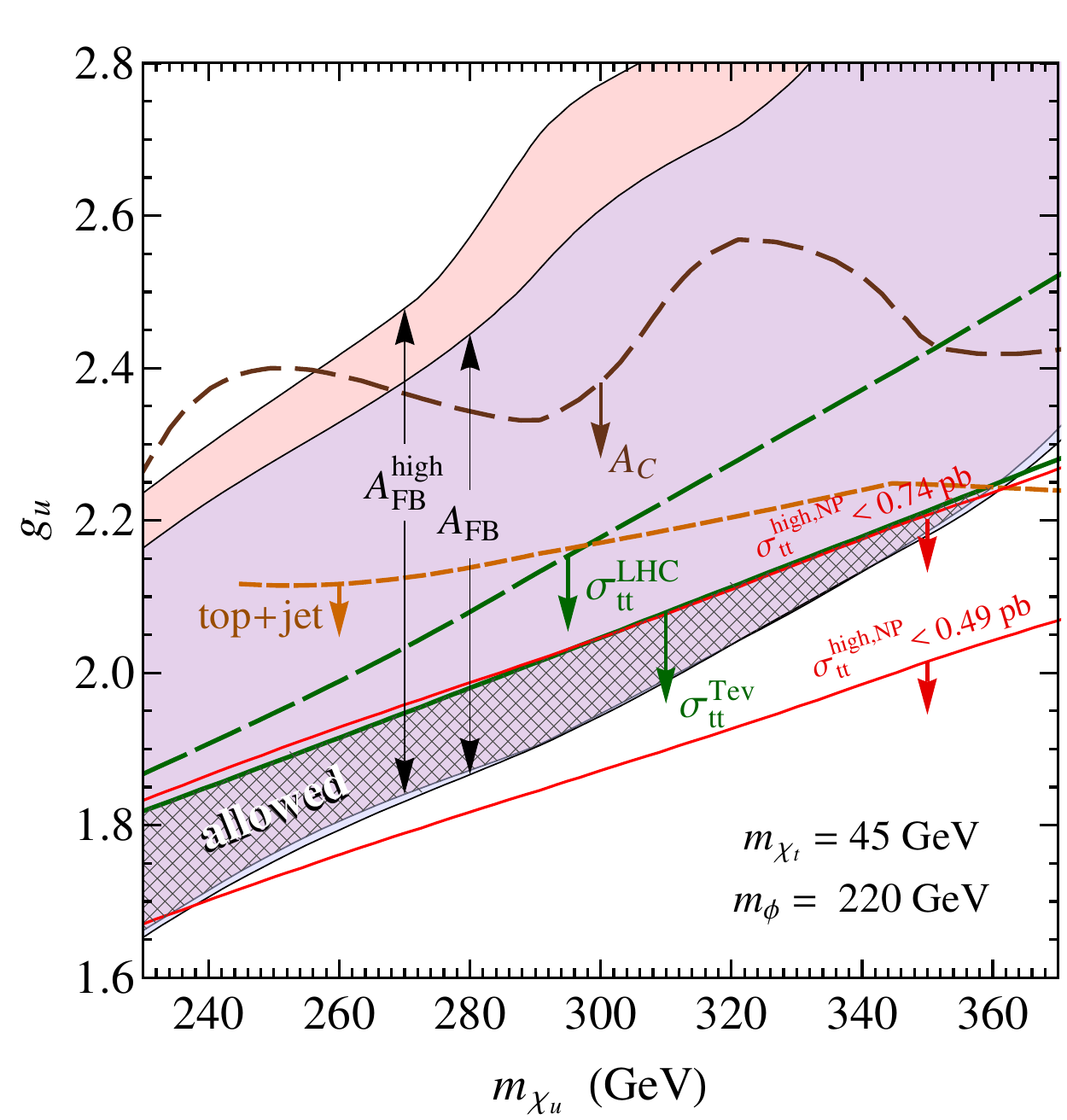}
\label{fig:chit45}
}
\subfigure[]{
\includegraphics[width=0.47\textwidth]{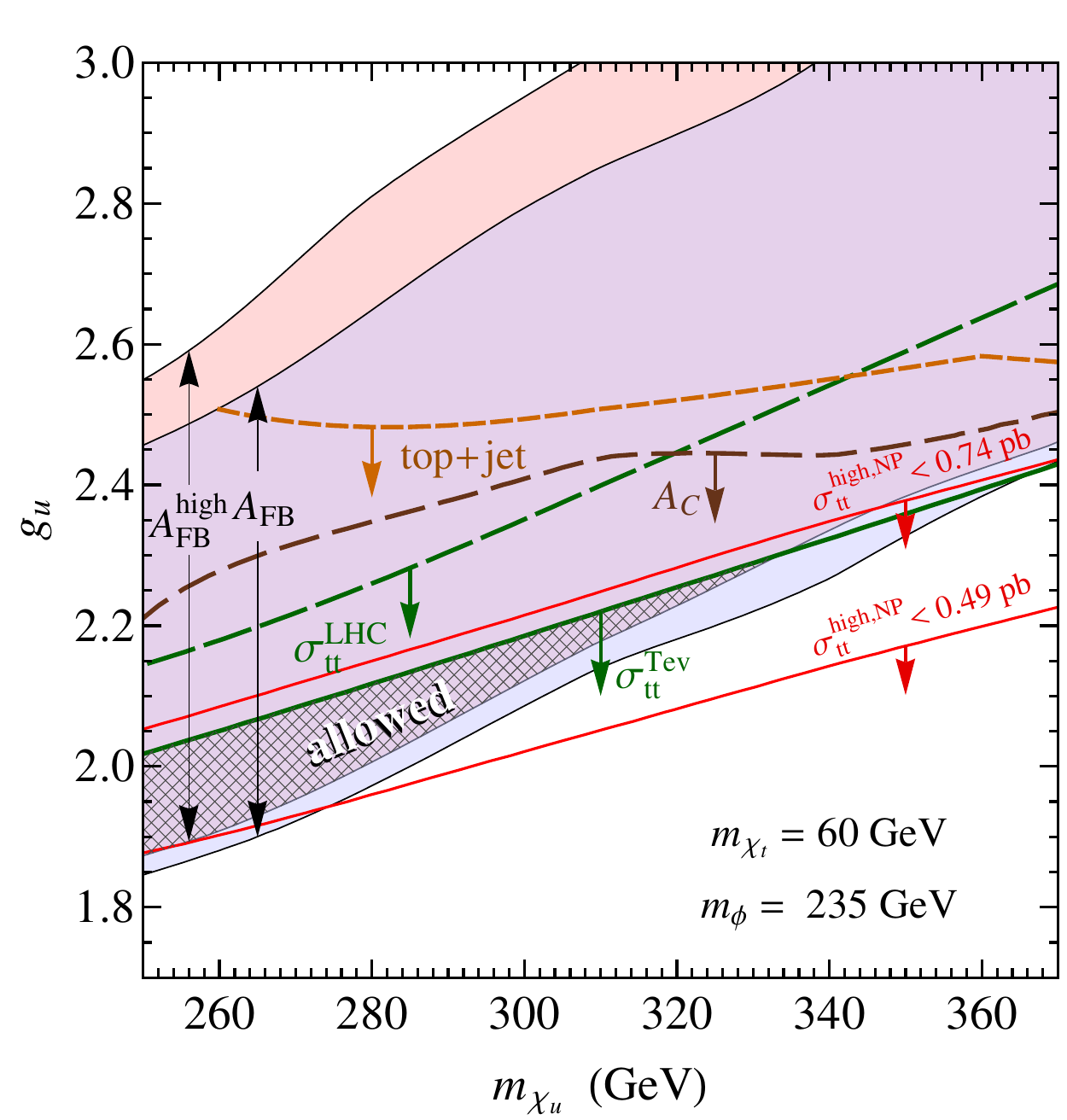}
\label{fig:chit60}
}
\subfigure[]{
\includegraphics[width=0.47\textwidth]{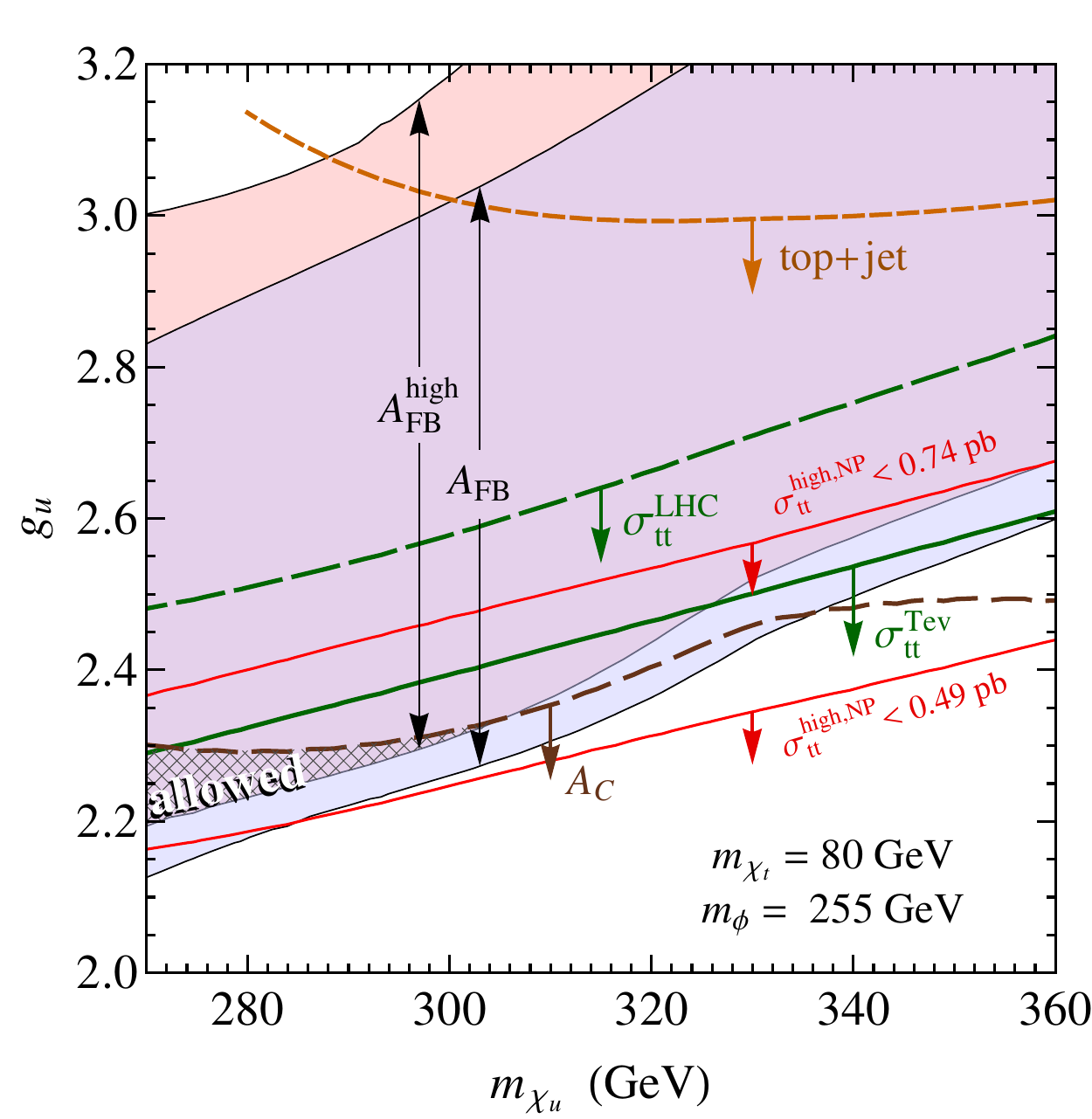}
\label{fig:chit80}
}
\caption{Allowed 1$\sigma$ region that fits top observables at Tevatron and LHC in the ($m_{\chi_{t}},g_{u}$) plane (a), and the ($m_{\chi_{u}},g_{u}$) plane with $m_{\chi_{t}}=45, \, 60, \, 80~\gev$, respectively (b,c,d).  Inclusive $\AFB$ ($\AFB^{\rm high}$) is consistent with the blue (red) region, with the overlap of both shown in purple.  Remaining Tevatron (LHC) observables are shown by the solid (dashed) lines, with allowed regions indicated by arrows.  The hatched region is consistent with all observables at $1\sigma$ except $\sigma_{t \bar t}^{\rm high}$.}  
\label{fig:topfigs}   
\end{figure}

Our main results are presented in Fig.~\ref{fig:topfigs}.  In general, the collider phenomenology of our model depends on five parameters $(g_u, g_t, m_{\chi_u}, m_{\chi_t}, m_\phi)$, and the different panels in Fig.~\ref{fig:topfigs} illustrate different slices through this parameter space.  In panel (a), we show our results in the ($m_{\chi_{t}}, g_{u}$) plane, assuming $m_\phi = m_{\chi_t} + 175$ GeV and $m_{\chi_u} = m_{\phi} + 10$ GeV.  In panels (b$-$d), we show our results in the ($m_{\chi_{u}}, g_{u}$) plane, for specific choices of $m_{\chi_t}$.  We assume $g_t = g_u$ for simplicity; since $g_t$ affects only the total width of $\phi$, this choice is of minor importance.\footnote{In Sec.~\ref{sec:det}, we show that relic abundance considerations fix $g_t$ as a function of $m_{\chi_t}$.}  

The different top constraints are shown as follows:
\begin{itemize}
\item The shaded red band is consistent with $\AFB^{\rm high} = (29.5 \pm 6.6)\%$, while the shaded blue region is consistent with inclusive $\AFB =  (17.4 \pm 3.7)\%$, taking an error-weighted average of Eq.~\eqref{AFBexpt}.  The central shaded purple band shows the overlap consistent with both.  To be conservative, we consider that $A_{FB}^{\rm SM}$ and $A_{FB}^{\rm high,SM}$ are at the upper limit of their $1\sigma$ allowed values.
\item The solid lines indicate Tevatron constraints.  The solid green line shows the $1\sigma$ region for the inclusive $t \bar t $ cross section $\sigma_{t \bar t}^{\rm TeV}$. The red contours indicate $\sigma_{t \bar t}^{\rm high,NP}$ at the level of $0.49$ and $0.74$ pb, corresponding to a $1.5\sigma$ and $2 \sigma$ enhancement over the difference of observed and SM values ($\sigma_{t \bar t}^{\rm high,NP}= - 0.25 \pm 0.49$ pb).  The arrows denote the allowed region.
\item The dashed lines indicate LHC constraints at 7 TeV on the inclusive $t \bar t$ cross section $\sigma_{t\bar t}^{\rm LHC}$ (green), top+jet resonance searches (orange), and $A_C = (0.1 \pm 1.4)\%$ (brown), taking an error-weighted average of Eq.~\eqref{ACexpt}.  The arrows denote the allowed region.
\end{itemize}
The hatched region is consistent with all aforementioned measurements at $1\sigma$ except for tension with $\sigma_{t\bar t}^{\rm high}$.    Taken at face value, the preferred (hatched) region for TFDM exhibits some tension at the level of $1.5-2 \sigma$ with $\sigma_{t \bar t}^{\rm high}$.  However, reconstruction and acceptance efficiencies, as well as higher-order corrections, which we have not addressed, may be important.

Fig.~\ref{fig:chitrange} shows the allowed DM mass and coupling that is consistent with Tevatron and LHC top data. A DM mass $m_{\chi_t} \sim 20-90$ GeV is allowed, with coupling $g_u \sim 1.4 - 2.2$.  Tevatron $t\bar t$ measurements provide the strongest constraints, while $m_{\chi_t} > 90$ GeV is excluded by $A_C$ and $m_{\chi_t} < 20$ GeV is excluded by $\sigma_{t \bar t}^{\rm LHC}$.  It is noteworthy that relic density constraints also point to a similar DM mass range, discussed in Sec.~\ref{sec:det}.

Figs.~\ref{fig:chit45}, \ref{fig:chit60}, \ref{fig:chit80} show how the allowed parameter region depends on $m_{\chi_u}$, for $m_{\chi_{t}} = 45, 60, 80~\gev$, respectively.  At larger $m_{\chi_u}$, the parameter region becomes more restricted by Tevatron $t\bar t$ measurements.  Improvements in sensitivity in top+jet resonance searches may play an important role in exploring TFDM parameter space.

\section{Relic density, direct detection, and other constraints}
\label{sec:det}

\begin{figure}[t]
\begin{center}    
\includegraphics[scale=1]{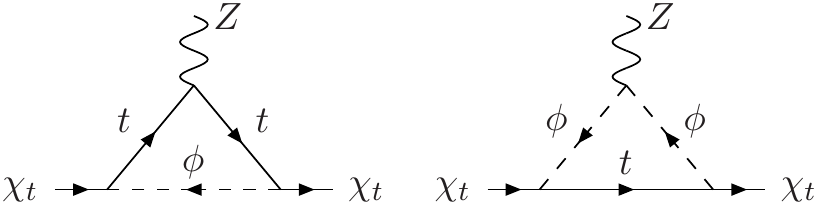}
\caption{Feynman diagrams for DM coupling to the $Z$ boson.}  
\label{Zfig}   
\end{center}  
\end{figure}

Top-flavored DM, by virtue of its large coupling to top, naturally acquires a significant one-loop coupling to the $Z$ boson.  This coupling sets the DM relic density and also generates observable signals in direct detection experiments.  From the diagrams shown in Fig.~\ref{Zfig}, the resulting effective interaction is
\beq \label{effZ}
\mathscr{L}_{\rm int} = a_Z \frac{g_2}{c_W}  \bar{\chi}_t \gamma^\mu P_L \chi_t Z_\mu \, , 
\eeq
where $g_2$ is the $SU(2)_L$ gauge coupling, $c_W \equiv \cos\theta_W$ is the cosine of the weak mixing angle, and $a_Z$ is the new physics coefficient, given by
\beq
a_Z = \frac{3 g_t^2}{32 \pi^2} \frac{m_t^2}{m_\phi^2} \mathcal{F}_Z(m_t^2/m_\phi^2) \, , \quad \mathcal{F}_Z(x) \equiv \frac{x-1-\log x}{(1-x)^2} \, .
\eeq
Although Eq.~\eqref{effZ} corresponds to a dimension-six operator $\bar{\chi}_t \gamma^\mu P_L \chi_t H^\dagger D_\mu H$, the mass scale suppression $1/\Lambda^2$ is compensated by the Higgs vacuum expectation value $\langle H \rangle^2$.  Within Fig.~\ref{Zfig}, $\langle H \rangle$ manifests as the mass of the top quark in the loop.  If the new physics scale is comparable to $m_t$, as we have assumed, there is no mass scale suppression of the DM coupling the $Z$, giving $a_Z \sim 0.01 g_t^2$.

In the early Universe, DM annihilates to SM fermions $f \bar f$ through the $Z$ boson.  The total annihilation cross section is
\beq \label{swaveann}
(\sigma v )_{\rm an} = \frac{ g_2^4 a_Z^2 m_{\chi_t}^2}{8\pi c_W^4 \big( (s - m_Z^2)^2 + m_Z^2 \Gamma_Z^2\big)} \times \sum_{f} N_c^f \big( T_{3f}^2 - 2 T_{3f} Q_f s_W^2 + 2 Q_f^2 s_W^4 \big) \, ,
\eeq
working at leading order ($s$-wave) in the relative velocity $v$, except we keep the full $v$-dependence of the propagator to properly treat resonant annihilation~\cite{Griest:1990kh}.
In Eq.~\eqref{swaveann}, $s \approx 4m_{\chi_t}^2 (1 + v^2/4)$ is the center-of-mass energy, $s_W \equiv \sin\theta_W$, and $N_c^f$, $Q_f$, and $T_{3f}$ denote the number of colors, electric charge, and weak isospin, respectively, summed over all SM fermions $f$ excluding the top, which is not kinematically allowed. 

For a Dirac particle, the required value is $\langle \sigma v \rangle_{\rm an} \approx 6 \times 10^{-26} \, \cms$ to achieve the observed DM relic abundance during freeze-out.\footnote{To clarify our notation, we note that $\langle \sigma v \rangle_{\rm an}$ is the thermal average of $(\sigma v)_{\rm an}$.}  Although $\langle\sigma v \rangle_{\rm an}$ is suppressed by the loop suppression of $a_Z$, it is enhanced by resonant annihilation for $m_{\chi_t} \sim m_Z/2$, by the large number of SM fermion final states, and by a large coupling $g_t \gtrsim 1$.  Following ref.~\cite{Griest:1990kh}, the DM relic abundance is
\be
\Omega_{\rm dm} h^2 \approx \frac{ 1.03 \times 10^{9} \, {\rm GeV}^{-1}}{g_{*s}/g_*^{1/2} \, m_{\rm pl} \, J(x_{\rm fo})} \; ,
\ee
where $x_{\rm fo} \equiv m_{\chi_t}/T_{\rm fo}$ ($T_{\rm fo}$ is the freeze-out temperature) is given by the solution to the equation $x_{\rm fo} = \log({0.038 g_{\chi_t}  m_{\rm pl}\, \langle \sigma v \rangle_{\rm an} }/{\sqrt{g_* x_{\rm fo}} })$,
with $g_{\chi_t} = 2$.  Post-freeze-out annihilation is parametrized by the integral
\be \label{Jfactor}
J(x_{\rm fo}) = \frac{1}{2} \int^{\infty}_{x_{\rm fo}} \frac{dx}{x^2} \, \langle \sigma v \rangle_{\rm an} \, , \quad \langle \sigma v \rangle_{\rm an} = \sqrt{\frac{x^3}{4 \pi}} \int^\infty_0 dv \, v^2 (\sigma v)_{\rm an} \, e^{-xv^2/4} \, ,
\ee
where the $1/2$ arises for Dirac fermions.  We evaluate these integrals numerically, which is important on-resonance ($m_{\chi_t} \approx m_Z/2$), where the $s$-wave approximation breaks down.

Next, we discuss several experimental constraints.  If DM is lighter than $m_Z/2$, decays $Z \to \chi_t \bar{\chi}_t$ are allowed, contributing to the invisible $Z$ width.  The partial width is
\beq
\Gamma(Z \to \chi_t \bar \chi_t) = \frac{ g_2^2 a_Z^2 m_Z}{96 \pi c_W^2} \,  \big( 1 - m_{\chi_t}^2/m_Z^2\big) \sqrt{ 1 - 4m_{\chi_t}^2/m_Z^2}  \approx 1.6 \; {\rm MeV} \times \left(\frac{a_Z}{0.1} \right)^2 \, .
\eeq
The measured invisible $Z$ width is consistent with the SM, $\Gamma(Z\to {\rm inv})_{\rm expt} - \Gamma(Z\to {\rm inv})_{\rm SM} = - 1.7 \pm 1.5$ MeV~\cite{Beringer:1900zz}, and any new physics contribution must be no greater than $\mathcal{O}({\rm MeV})$.  We also note that TFDM does not contribute to highly-constraining parity-violating observables at one-loop, unlike standard $t$-channel models~\cite{Gresham:2011dg}.

Direct detection experiments, which search for nuclear recoils from DM in the local halo, have important implications for top-flavored DM.  In particular, the DAMA/LIBRA~\cite{Bernabei:2008yi}, CoGeNT~\cite{Aalseth:2010vx}, and CRESST-II~\cite{Angloher:2011uu} experiments have found evidence for a signal from $\mathcal{O}(10 \, {\rm GeV})$ mass DM; however, taken at face value, these observations appear at odds with null results from the XENON100~\cite{Aprile:2012nq}, XENON10~\cite{Angle:2011th}, and CDMS~\cite{Ahmed:2009zw,Ahmed:2010wy} experiments.  Reconciling these different results in an area of active debate, with appeals to astrophysical assumptions, experimental systematic effects, or DM model-dependent assumptions (see~\cite{Chang:2010yk,Kelso:2011gd,Frandsen:2011gi,Arina:2012dr} and refs.~therein).  

\begin{figure}[t]
\begin{center}    
\includegraphics[scale=.9]{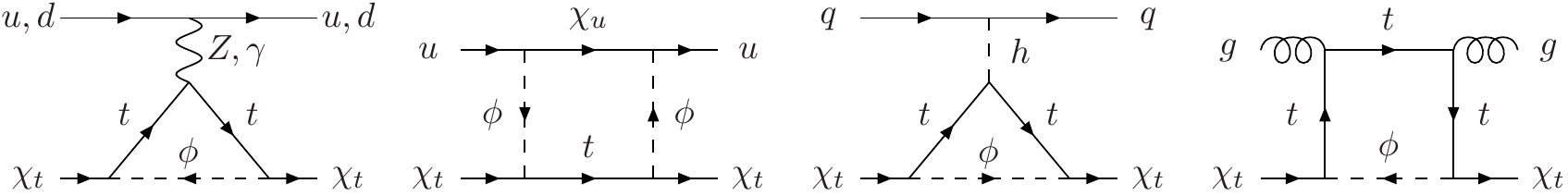}
\caption{Feynman diagrams contributing to the SI DM-nucleon cross section relevant for direct detection experiments. For each case we show only one representative diagram, while we include all possible diagrams in our analysis.}  
\label{DDfigs}   
\end{center}  
\end{figure}

Although $\chi_t$ has no tree-level couplings to nucleons, couplings arise at one-loop order, shown in Fig.~\ref{DDfigs}.  
For a given nucleus $N$, the SI DM-nucleon scattering cross section is (see e.g.~\cite{Jungman:1995df,Beltran:2008xg})
\be
\sigma_{N}^{\rm SI} = \frac{\mu_N^2}{\pi} \, \big( Z f_p + (A-Z) f_n \big)^2 
\ee
where $(Z,A)$ are the proton and atomic numbers, and $\mu_N \equiv m_N m_{\chi_t}/(m_N + m_{\chi_t})$ is the reduced mass.  The proton/neutron scattering amplitude is $f_{p,n} = f_{p,n}^Z + f_{p,n}^{\rm box} + f_{p,n}^h + f_{p,n}^g$, where the different terms delineate each contribution in Fig.~\ref{DDfigs}.  The $Z$ and box terms  provide the leading contributions to $\sigma_N$ through an effective vector-vector operator $\bar q \gamma^\mu q \bar\chi_t \gamma_\mu \chi_t$, with $q = u,d$.
The $Z$ vertex contribution is 
\be
f_n^Z = \frac{G_F a_Z}{\sqrt{2}} \, , \quad f_p^Z = -(1-4 s_W^2) \frac{G_F a_Z}{\sqrt{2}} \; ,
\ee
and the box contribution is
\be
f_p^{\rm box} =  \frac{ a_{\rm box}}{2} \, , \quad f_n^{\rm box} = \frac{ a_{\rm box}}{4} \, , \quad  \quad a_{\rm box} = \frac{g_u^2 g_t^2 }{16\pi^2 m_\phi^2 } \mathcal{F}_{\rm box}(m_t^2/m_\phi^2,m_{\chi_u}^2/m_\phi^2)
\ee
with loop function
\be
\mathcal{F}_{\rm box}(x,y) \equiv \frac{ x^2(y-1)^2 \log x - y^2 (x-1)^2 \log y + (x-1)(y-1)(x-y)}{4(x-1)^2(y-1)^2(x-y)} \, .
\ee
The Higgs-exchange amplitude is
\be
f_{p,n}^{h} = \frac{G_F g_t^2 m_{p,n} m_{\chi_t}}{16\pi^2 \sqrt{2} m_h^2} \left( 1 - \frac{7}{9} f_{TG}^{(p,n)} \right) \, \mathcal{F}_h(m_t^2/m_\phi^2) \, , \quad \mathcal{F}_h(x) \equiv \frac{ x^3 - 2x^2 \log x - x}{(x-1)^3} \, ,
\ee
 where $f_{TG}^{(p,n)} \approx 0.83$ is a hadronic matrix element~\cite{Ellis:2000ds}.  For the gluon amplitude, using results from ref.~\cite{Hisano:2010ct}, we obtain (for the contribution denoted therein as $f_G$)
\be
f_{p,n}^g = - \frac{g_t^2}{36} m_{\chi_t} m_{p,n} f^{(p,n)}_{TG} \left( f_+^s + f_+^l \right) \, ,
\ee
with loop functions $f^{s,l}_+$ defined in ref.~\cite{Hisano:2010ct} (with quark mass $m_t$ and ``squark'' mass $m_\phi$).  The Higgs-exchange and gluon amplitudes are quantitatively much smaller than the $Z$ and box terms, providing $\mathcal{O}(0.1\%)$ corrections to $f_{p,n}$.  The $\gamma$ exchange contribution corresponds to a magnetic dipole interaction, with a $\chi_t$ dipole moment
\be
\mu_{\chi_t} \approx \frac{e g_t^2 m_{\chi_t}}{32 \pi^2 m_{\phi}^2} \mathcal{F}_\gamma(m_t^2/m_\phi^2) \, , \quad \mathcal{F}_\gamma(x) \equiv \frac{1+2 x \log x-x^2}{(1-x)^3} \; .
\ee
For 10 GeV DM, a magnetic dipole moment at the level of $10^{-18} \, e \, {\rm cm}$ can potentially alleviate tensions in direct detection results due to the momentum dependence of the cross section~\cite{DelNobile:2012tx}.  Here, we have $\mu_{\chi_t} \approx 7 \times 10^{-21} g_t^2 \, e \, {\rm cm}$ for $m_\phi \approx m_t$, which is too small unless $g_t$ in nonperturbatively large, in which case $f_n^Z$ is similarly enhanced.  Therefore, we neglect $\mu_{\chi_t}$ in our analysis.

\begin{figure}[t]
\begin{center}    
\includegraphics[scale=.64]{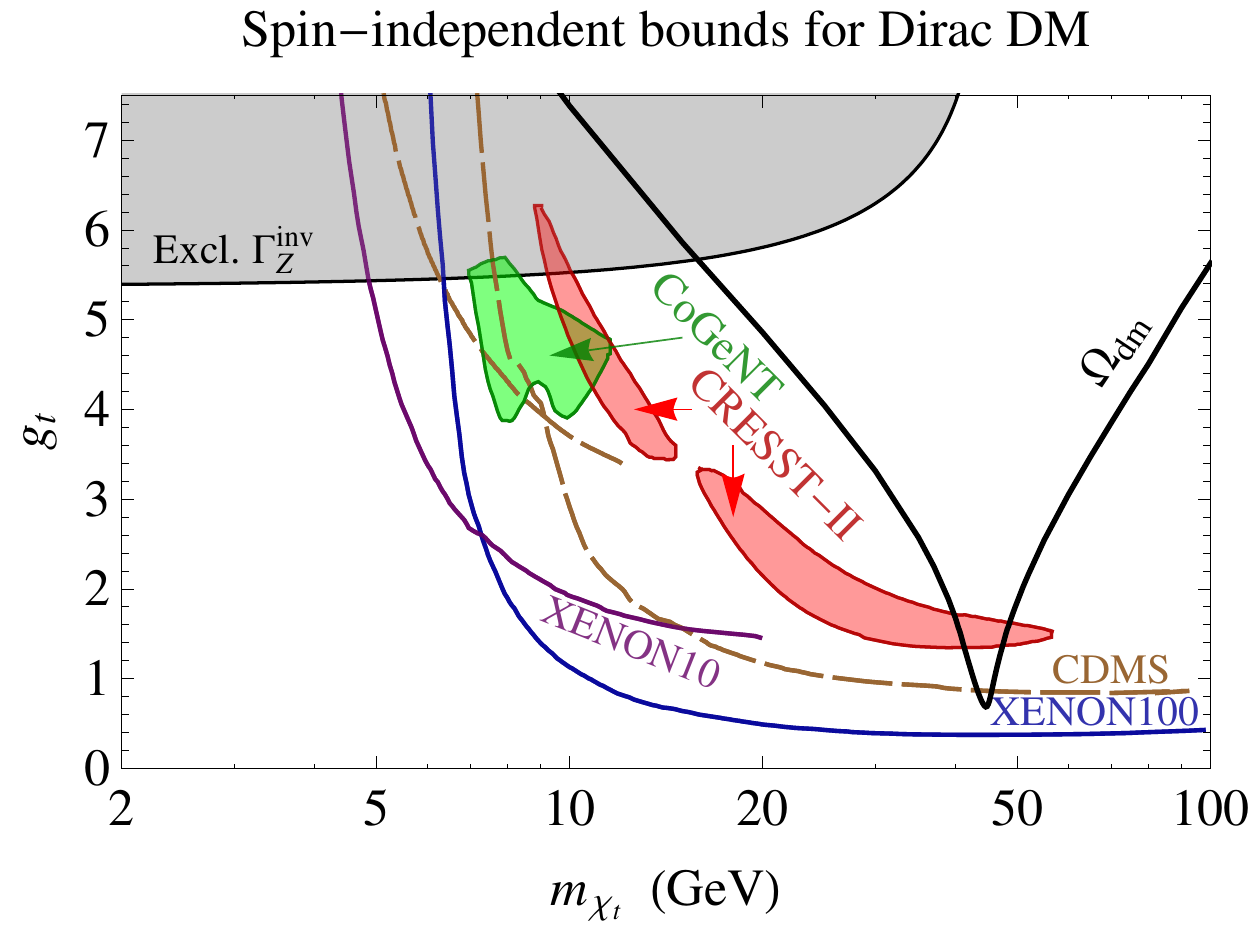} \includegraphics[scale=.64]{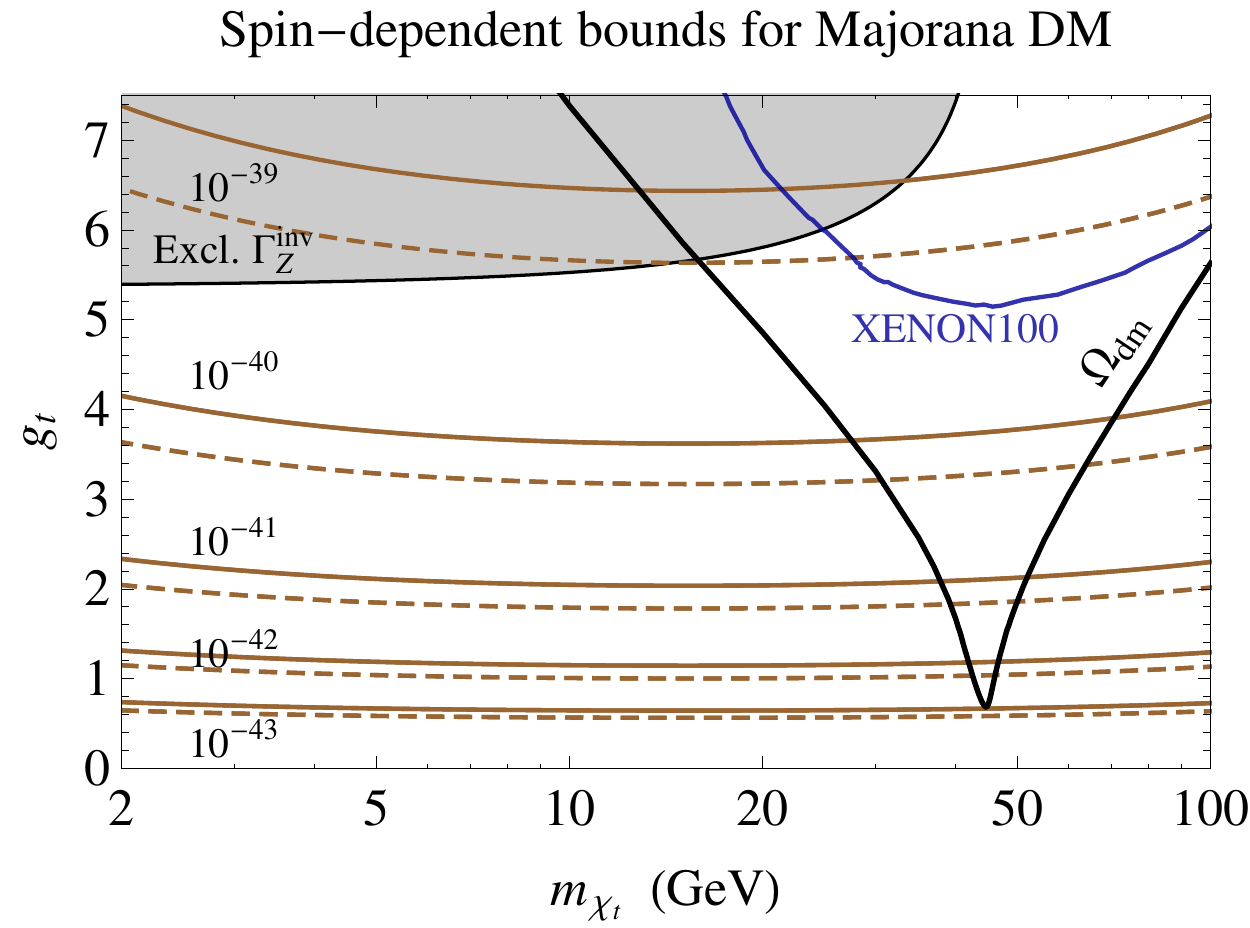}
\caption{Left: For Dirac DM, we show exclusion curves for SI scattering from XENON100~\cite{Aprile:2012nq} (blue), XENON10~\cite{Angle:2011th} (purple), and CDMS~\cite{Ahmed:2009zw,Ahmed:2010wy} (brown dashed), while enclosed regions show signal regions from CoGeNT~\cite{Aalseth:2010vx} (green) and CRESST-II~\cite{Angloher:2011uu} (red).  Right: For Majorana DM, we show exclusion limit on SD scattering from XENON100~\cite{Garny:2012it}, as well as the prospective reach for limits for SD scattering on neutrons (solid brown) and protons (dashed brown).  Gray region is excluded by $\Gamma(Z \to {\rm inv})$.  Black line gives the DM relic density $\Omega_{\rm dm} h^2= 0.11$.}  
\label{DDplot}   
\end{center}  
\end{figure}

In Fig.~\ref{DDplot}, we show how constraints from direct detection experiments map onto top-flavored DM parameter space, as a function of $(g_t, m_{\chi_t})$.  For definiteness, we have fixed $m_\phi - m_t - m_{\chi_t} = 2$ GeV; we have also taken $g_u = 2$ and $m_{\chi_u}= 400$ GeV for evaluating $f^{\rm box}_{p,n}$, although these parameter choices are of minor importance since the box term provides only a subdominant contribution to the scattering cross section.  The solid black line (labeled $\Omega_{\rm dm}$) shows the parameters required for $\Omega_{\rm dm} h^2= 0.11$~\cite{Beringer:1900zz}, while the gray region is excluded for $\Gamma(Z \to \chi_t \bar \chi_t) < 3$ MeV.  Provided $\chi_t$ is a Dirac state, constraints on SI scattering provide the strongest limits, due to the sizable one-loop coupling to the $Z$.  The left panel of Fig.~\ref{DDplot} shows that the entire parameter space that gives the correct relic density is excluded by XENON100, as well as other measurements.  These constraints require $m_{\chi_t} \lesssim 10$ GeV or $g_t \lesssim 0.1$.  In this case, annihilation through the $Z$ is insufficient to achieve the correct relic density, and additional channels would be required.

The stringent limits from SI scattering can be evaded if DM is Majorana.  We suppose that the Dirac state $\chi_t$ acquires a flavor-breaking Majorana mass term, given in Eq.~\eqref{majoranamass}, that splits $\chi_t$ into two Majorana states $\chi_{1,2}$, with only the lighter state $\chi_1$ populated in the Universe today.  In terms of $\chi_{1,2}$, Eq.~\eqref{effZ} becomes
\be
\mathscr{L}_{\rm int} = a_Z \frac{g_2}{2 c_W} \left( \bar{\chi}_2 \gamma^\mu \chi_1 + \bar{\chi}_1 \gamma^\mu \chi_2 + \bar{\chi}_1 \gamma^\mu \gamma_5 \chi_1 + \bar{\chi}_2 \gamma^\mu \gamma_5 \chi_2 \right) Z_\mu \; .
\ee
The vector interaction becomes inelastic.   If the mass splitting $\Delta m$ is larger than $\mathcal{O}(100 \, {\rm keV})$, $\chi_1$ has insufficient kinetic energy to access the excited state $\chi_2$~\cite{TuckerSmith:2001hy}.  On the other hand, the axial vector interaction remains elastic, contributing to the SD DM-nucleon scattering cross section.  Following ref.~\cite{Jungman:1995df}, the SD cross section on the neutron or proton is given by
\be
\sigma_{n,p}^{\rm SD} = \frac{12}{\pi} \mu_{p,n}^2 \Big( d_u \Delta_u^{(p,n)} + d_d \Delta_d^{(p,n)} + d_s \Delta_s^{(p,n)}\Big)^2
\ee
for axial-vector interaction $\mathscr{L}_{\rm eff} = d_q \bar q \gamma^\mu \gamma_5 q \bar \chi_1 \gamma_\mu \gamma_5 \chi_1$.  The coefficients $d_q$ are given by
\be
d_u = - \frac{G_F a_Z}{2\sqrt 2}  - \frac{a_{\rm box}}{8} \, , \quad d_d = d_s = \frac{G_F a_Z}{2\sqrt 2}  \, ,
\ee
due to $Z$ exchange and box contributions ($d_u$ only).  We take $\Delta_u^{(p)} = \Delta_d^{(n)} = 0.78$, $\Delta_d^{(p)} = \Delta_u^{(n)} = -0.48$, $\Delta_s^{(p)} = \Delta_s^{(n)} = -0.15$~\cite{Ellis:2000ds}.

Fig.~\ref{DDplot} (right) shows how constraints from SD DM-nucleon scattering map onto TFDM parameter space.  XENON100 provides the strongest limits on TFDM scattering on the neutron~\cite{:2013uj}, shown by the blue line (see also ref.~\cite{Garny:2012it}).\footnote{Constraints from SI interactions (via Higgs and gluon amplittudes) are much weaker and lie above the vertical scale of the plot.}  The combination of the constraints from XENON100 SD scattering and $\Gamma(Z \to {\rm inv})$ exclude $g_t \gtrsim 5$, while a large parameter region remains viable.  The brown solid (dashed) contours indicate the SD cross section (in ${\rm cm}^2$ units) for DM scattering on the neutron (proton), showing how future searches will impact TFDM.  The relic abundance constraint provides a lower bound $\sigma_{n,p}^{\rm SD} \gtrsim 10^{-43} \, {\rm cm}^2$, which is saturated for $m_{\chi_t} \approx 45$ GeV.  Future experiments, such as XENON1T~\cite{Aprile:2012zx}, at the level of $\sigma_n^{\rm SD} \sim 10^{-41} - 10^{-42} \; {\rm cm}^2$ offer the potential to explore nearly all of TFDM parameter space, except for very near the $Z$ pole.

The Majorana mass splitting $\Delta m$ can also play an important role for DM annihilation.  The annihilation cross section for $\chi_1 \chi_1 \to f \bar f$ is $p$-wave or chirality-suppressed, given by
\be
(\sigma v) =\frac{ g_2^4 a_Z^2 }{8\pi c_W^4 \big( (s - m_Z^2)^2 + m_Z^2 \Gamma_Z^2\big)} \sum_f N_c^f \left(  T_{3f}^2 m_f^2  + \frac{2}{3} m_1^2 v^2 (T_{3f}^2 - 2 T_{3f} Q_f s_W^2 + 2 Q_f^2 s_W^4 ) \right) ,
\ee
while $\chi_1 \chi_2$ annihilation has an unsuppressed $s$-wave component given by Eq.~\eqref{swaveann}.  This implies that DM annihilation is suppressed in the halo today (since only $\chi_1$ is populated), and thus indirect detection signals are quenched.  However, if $\Delta m \lesssim T_{\rm fo} \sim {\rm  GeV}$, then $\chi_2$ is populated during freeze-out, and the correct relic density is obtained by $\chi_1 \chi_2 \to f \bar f$ coannihilation, as in the Dirac case.  Thus, the relic density line (black) in Fig.~\ref{DDplot} (right) provides a lower bound on $g_t$.   The bound is saturated for $\Delta m \ll T_{\rm fo}$, while larger $\Delta m \sim T_{\rm fo}$ requires larger values of $g_t$ since the $\chi_1 \chi_2$ annihilation rate has a Boltzmann suppression $\exp(-\Delta m/T_{\rm fo})$.

Although we have thus far assumed a standard (symmetric) freeze-out cosmology, it is also possible that $\chi_t$ might be asymmetric DM (see~\cite{Kaplan:2009ag,Davoudiasl:2012uw} and refs.~therein).  Within extra-dimensional realizations, TFDM carries a generalized baryon number, and therefore the dark sector may have a baryon asymmetry similar to visible sector (in fact, they may be naturally related)~\cite{Agashe:2004ci,Agashe:2004bm}.  To the extent that $\Delta m \approx 0$ and $\chi_t$-$\bar \chi_t$ annihilation is efficient, the relic density is set by the initial dark baryon asymmetry.  The natural mass scale for $\chi_t$ is $(\Omega_{\rm dm}/\Omega_{\rm b}) \times m_p \sim 5$ GeV.  While there is no Majorana splitting to evade SI limits, asymmetric DM naturally favors the low mass window where SI limits are weakened.\footnote{Introducing a Majorana splitting as large as $\mathcal{O}(100 \, {\rm keV})$ leads to $\chi_t$-$\bar\chi_t$ oscillations that wash out any initial asymmetry, leading to the usual symmetric freeze-out scenario.}  On the other hand, efficient symmetric annihilation requires $\langle \sigma v \rangle_{\rm an} \gtrsim 6 \times 10^{-26} \; \cms$.  Since annihilation through the $Z$ is insufficent in the low mass window, additional channels are required in this case.

\section{Conclusions}
\label{sec:conclude}

The top forward-backward asymmetry remains a persistent anomaly from the Tevatron.  Many new physics models have been proposed, with new degrees of freedom introduced mainly to address this anomaly.  Here, we have shown that possible new physics in the top sector may be connected to another long-standing particle physics puzzle -- the nature of dark matter. We have considered a simple phenomenogical model where dark matter $\chi_t$ carries top flavor and couples to the SM via the top quark within the framework of MFV.  In particular, we find that top physics and dark matter constraints \emph{independently} point toward the same mass range for DM: $m_{\chi_{t}} \sim 20-90~\gev.$

For collider phenomenology, the scalar mediator $\phi$ plays a key role, with an $\mathcal{O}(1)$ forward-backward asymmetry in on-shell $\phi \phi^*$ production transferred into a $t \bar t$ asymmetry through decays $\phi \to t \bar \chi_t$, as in ref.~\cite{Isidori:2011dp}.  Missing energy signatures in $t \bar t$ events are avoided for $m_\phi -m_t - m_{\chi_t} \approx 0$, and several constraints on the model of ref.~\cite{Isidori:2011dp} are evaded due to the additional flavor structure of our model.  We have shown that top-flavored DM generates a sizable top asymmetry and is consistent with other Tevatron and LHC observables for $m_{\chi_t} \sim 20 - 90$ GeV and for a perturbative range of couplings, albeit with some tension the high invariant mass $t \bar t$ cross section measured by the Tevatron.  Future top+jet resonance analyses should be able to explore the parameter space of the model. 

The DM relic density and direct detection signals are determined by an effective coupling of DM to the $Z$ boson.  This coupling arises at one-loop and is enhanced by the mass of the top appearing in the loop. The DM freeze-out abundance is set by annihilation $\chi_t \bar\chi_t \to f\bar f$ through the $Z$ vector coupling.  We have shown that the correct relic density can be achieved for DM masses prefered by collider constraints (especially near the $Z$ pole).  If DM has a small (flavor-breaking) Majorana mass splitting $100 \; {\rm keV} \lesssim \Delta m \lesssim {\rm GeV}$, direct and indirect detection processes are governed by the $Z$ axial-vector coupling.  Stringent SI limits are evaded, and the most promising direct detection signals are for SD scattering (with a lower bound $\sigma_{p,n}^{\rm SD} \gtrsim 10^{-43} \; {\rm cm}^2$ at $m_{\chi_t} \approx m_Z/2$).  Annihilation in the halo today is $p$-wave or chirality suppressed. Our model illustrates that DM annihilation through the $Z$ boson remains a viable mechanism for fixing the relic density, and future SD experiments are crucially important for exploring this possibility.

\acknowledgements  
 
We thank J.~Alwall, K.~Blum, J.~Kamenik, I.-W.~Kim, S.~McDermott, D.~Morrissey, N.~Weiner, P.~Winslow for helpful discussions.  We acknowledge the use of \texttt{DMTools} to obtain direct detection experimental limits shown in Fig.~\ref{DDplot}.  AK is supported by the National Science and Engineering Research Council of Canada.  ST is supported by the Department of Energy under contract de-sc0007859.  As this work neared completion, Ref.~\cite{Haisch:2013uaa} also noted the importance of direct detection signals arising radiatively through the top.

\bibliography{afbbiblio}

\end{document}